\documentclass[envcountsect, envcountsame]{llncs}

\usepackage{textcomp}
\usepackage{amsmath}
\usepackage{amssymb}
\usepackage{graphics}
\usepackage{psfrag}

\newcommand{\E}{\ensuremath{\mathbf{E}}}

\newcommand{\vbl}{\ensuremath{{\rm vbl}}}
\newcommand{\ex}{\ensuremath{{\rm ex}}}
\renewcommand{\P}{\ensuremath{{\mathcal{P}}}}

\newcommand{\Fk}{\ensuremath{\mathbb{F}_k}}

\begin{document}
\frontmatter
\pagestyle{headings}
\title{Unsatisfiable Linear $k$-CNFs Exist, for every $k$}

\author{Dominik Scheder}
\institute{Theoretical Computer Science, ETH Z\"urich\\
CH-8092 Z\"urich, Switzerland\\
\email{dscheder@inf.ethz.ch}}

\maketitle

\begin{abstract}
  We call a CNF formula {\em linear} if any two clauses have at most
  one variable in common.  Let Linear $k$-SAT be the problem of
  deciding whether a given linear $k$-CNF formula is satisfiable.
  Here, a $k$-CNF formula is a CNF formula in which every clause has
  size exactly $k$.  It was known that for $k \geq 3$, Linear $k$-SAT
  is NP-complete if and only if an unsatisfiable linear $k$-CNF
  formula exists, and that they do exist for $k \leq 4$. We prove that
  unsatisfiable linear $k$-CNF formulas exist for every $k$. Let
  $f(k)$ be the minimum number of clauses in an unsatisfiable linear
  $k$-CNF formula. We show that $f(k) \in \Omega(k2^k) \cap O(4^k
  k^4)$, i.e., minimum size unsatisfiable linear $k$-CNF formulas are
  significantly larger than minimum size unsatisfiable $k$-CNF
  formulas.  Finally, we prove that, surprisingly, linear $k$-CNF formulas do
  not allow for a larger fraction of clauses to be satisfied than
  general $k$-CNF formulas.
\end{abstract}

\section{Introduction}

A CNF formula $F$ (conjunctive normal form) over a variable set $V$ is
a set of clauses; a clause is a set of literals; a literal is either a
variable $x \in V$ or its negation $\bar{x}$. A CNF formula $F$, or
short, a CNF $F$, is called a $k$-CNF if $|C| = k$ for every $C \in
F$.  Define $\vbl(x)=\vbl(\bar{x}):=x$ for $x \in V$,
$\vbl(C):=\{\vbl(l) \ | \ l \in C\}$ and $\vbl(F) := \bigcup_{C\in F}
\vbl(C)$. For example, $\vbl(\{\bar{x},y,\bar{z}\}) = \{x,y,z\}$. A
(partial) assignment $\alpha$ is a (partial) function $V \rightarrow
\{0,1\}$. It can be extended to negated variables by $\alpha(\bar{x})
:= \neg \alpha(x)$.  A clause is {\em satisfied} by $\alpha$ if at
least one literal in it evaluates to $1$, and a formula is satisfied
if every clause is satisfied.  {\em Applying} a partial assignment
$\alpha$ means removing from $F$ every clause satisfied by $\alpha$,
and from the remaining clauses removing all literals evaluating to
$0$. The
resulting formula is denoted by $F^{[\alpha]}$.\\

Consider a set system $S$ of sets of cardinality $k$ over some ground
set $V$, i.e. a $k$-uniform hypergraph. We say $S$ is a $k$-set
system. We call $S$ {\em linear} if $|A \cap B| \leq 1$ for any $A, B
\in S, \ A \ne B$. We do not use any deep results from hypergraph
theory in this paper. Nevertheless, for definitions and basic terminology of
hypergraphs we refer the reader to~\cite{handbook} or~\cite{berge}.\\

A CNF $F$ is {\em linear} if $|\vbl(C) \cap \vbl(D)| \leq 1$ for all
clauses $C,D\in F, \ C \ne D$.  The set system $\{ \vbl(C),\ C \in
F\}$ is called the {\em skeleton} of $F$.  If $F$ is a $k$-CNF, then
its skeleton is a $k$-uniform hypergraph, which is linear if
$F$ is linear. Note that the converse does not hold in general:
The formula $\{\{x,y\},\{\bar{x},\bar{y}\}\}$ is not linear, but its skeleton
is $\{\{x,y\}\}$, thus linear.\\

\textbf{Examples:} The formula
$\{\{\bar{x}_1,x_2\},\{\bar{x}_2,x_3\},\{\bar{x}_3,x_4\},\{\bar{x}_4,x_1\}\}$
is a linear $2$-CNF, whereas 
$\{\{x_1,x_2,x_3\},\{\bar{x}_2\,x_3,x_4\}\}$ is not linear.\\

\subsection*{Previous Results}

Let $k$-SAT be the problem of deciding whether a given $k$-CNF is
satisfiable. It is well-known that $k$-SAT is NP-complete for $k \geq
3$. Define Linear $k$-SAT to be the corresponding decision problem for
linear $k$-CNFs. Porschen, Speckenmeyer and Randerath~\cite{porschen}
observed that Linear $k$-SAT is NP-complete if and only if there
exists an unsatisfiable linear $k$-CNF. They proved the existence of
unsatisfiable linear $k$-CNFs for $k=2,3$. In~\cite{porschen-neu},
Porschen, Speckenmeyer and Zhao prove existence for $k=4$.
Up to now, for $k\geq 5$ the question whether unsatisfiable $k$-CNFs exist
has been open.

\subsection*{Our Contribution}
We show that unsatisfiable linear $k$-CNFs exist for any $k$, hence
establishing NP-completeness of Linear $k$-SAT for all $k\geq 3$.
Further, let $f(k)$ denote the size of a smallest unsatisfiable linear
$k$-CNF.  We prove that $f(k) \in O(k^4 4^k)$ and, using the
Lov\'{a}sz Local Lemma, show that $f(k) \in \Omega(k2^k)$.  This is in
contrast to the general (non-linear) case, where
we know that unsatisfiable $k$-CNFs with $2^k$ clauses exist.\\

Having established $f(k) \in O(k^44^k)$, we are still looking for
explicit constructions of unsatisfiable linear $k$-CNFs. We give a
construction using $\leq t(k)$ clauses, for $t(0) := 1$ and $t(k+1) :=
t(k)2^{t(k)}$, i.e., a tower-like function. Compared to the gigantic
growth of $t(k)$, even $k^4 4^k$ seems very modest.

\section{Preliminaries}

Denote by $L(n,k)$ the maximum number of sets a linear $k$-set system
over $n$ elements can have. In this section, we give some bounds on
$L(n,k)$.  Everything in this section is standard graph and hypergraph
theory. The following upper bound is an easy observation.
See Theorem 3 in Chapter 1 of~\cite{berge} for example.

\begin{lemma}
\begin{displaymath}
L(n,k) \leq \frac{n(n-1)}{k(k-1)}
\end{displaymath}
\label{upper-bound-set-system}
\end{lemma}

\textbf{Proof.} Let $S$ be a linear $k$-system over $n$ elements.
There are ${n \choose 2}$ pairs of elements, and each $k$-set in $S$
contains ${k \choose 2}$ pairs. Since each pair is present in at most
one set, we obtain
$|S| \leq {n \choose 2}/{k \choose 2}$. $\hfill\Box$\\

If this upper bound is achieved, then every pair of elements occurs in
exactly one set, and the set system $S$ is also called a {\em Steiner
  system}. For existence of Steiner systems for specific values of $n$
and $k$ see for example~\cite{lindner}. At this point, we only give a
proof of existence of Steiner systems for $k$ being a prime power.

\begin{lemma}
  For every prime power $k$, there are infinitely many $n$ such that
$$
L(n,k) = \frac{n(n-1)}{k(k-1)}
$$
\end{lemma}

\textbf{Proof.} Let $k$ be any prime power, and let $\Fk$ be the
finite field of cardinality $k$. Let $\Fk^d$ be the $d$-dimensional
vector space over $\Fk$. It has $n = |\Fk^d| = k^d$ elements, called
{\em points}. For $x, y \in \Fk^d$ and $y \ne 0$, the set $\{x +
\lambda y \ | \ \lambda \in \Fk\}$ is called a {\em line}. A line
contains exactly $k$ points, and the vector space $\Fk^d$ has
$$
\frac{{n \choose 2}}{{k \choose 2}} = \frac{n(n-1)}{k(k-1)}
$$ 
lines: Every pair of distinct points $a$ and $b$ lies on exactly one
line, namely $\{a + \lambda (b-a) \big| \lambda \in \Fk\}$, and each
line can contains ${k \choose 2}$ pairs of distinct points. Note that
two lines intersect in at most one point.  Let $S$ be the $k$-set
system of all lines in $\Fk^d$. Then $S$ is a linear set system over
$n$ points, and $|S| = \frac{n(n-1)}{k(k-1)}$. $\hfill\Box$\\

If $k$ is not a prime power, we have the following weaker bound on $L(n,k)$. 

\begin{lemma} For any $n,k \in \mathbb{N}$,
\begin{displaymath}
L(n,k) \geq \frac{2n(n-1)}{k^2(k-1)^2}\ .
\end{displaymath}
\label{lower-bound-set-system}
\end{lemma}

\textbf{Proof.} Recall that any simple graph $G$ on $n$ vertices with
maximum degree $\Delta$ has an independent set $I \subseteq V$ with
$|I| \geq \frac{n}{\Delta+1}$.  This follows from a greedy construction: As
long as $G$ is not empty, pick a vertex and insert it into $I$. Remove
it and all its $\leq \Delta$ neighbors. In every
step, $\leq \Delta+1$ vertices are removed, hence we add  
at least $\frac{n}{\Delta+1}$ vertices to $I$.\\

For $n,k \in \mathbb{N}$, define a graph as follows: The vertices of
the graph are all ${n \choose k}$ $k$-sets over $n$ elements, and two
sets are connected by an edge if they share more than one element.
Each independent set of the graph corresponds to a linear $k$-set
systems over these $n$ elements. We estimate the maximum degree of
this graph. Let $s$ be a $k$-set. How many sets share two or more
elements with $s$?  There are $k \choose 2$ possibilities to fix $2$
elements to be included in the neighbor set $s'$, and ${n-2 \choose
  k-2}$ possibilities to choose the rest. Of course, this will
overcount the number of such sets. Hence there are at most ${k \choose
  2}{n-2 \choose k-2}$ sets sharing two or more elements with $s$.
Since $s$ itself is counted among those, we have $\Delta+1\leq
{k \choose 2}{n-2 \choose k-2}$. The graph itself has ${n \choose k}$
vertices, hence
$$
L(n,k) \geq \frac{{n \choose k}}{{k \choose 2}{n-2 \choose k-2}} \ ,
$$
and the lemma follows from a simple calculation.$\hfill \Box$\\

\section{Unsatisfiable $k$-CNFs formulas and NP-hardness}
In this section, we will prove existence of unsatisfiable $k$-CNFs for
any $k$, as well as proving some upper bounds on $|F|$, the number of
clauses in such a formula. Porschen, Speckenmeyer and
Randerath~\cite{porschen} already stated that for $k \geq 3$, Linear
$k$-SAT is NP-hard if there exists an unsatisfiable linear $k$-CNF.
To keep this paper self-contained, we include a proof of this result.

\begin{theorem}[Porschen, Speckenmeyer and Randerath~\cite{porschen}]
  For any $k\geq 3$, Linear $k$-SAT is NP-complete if there exists an
  unsatisfiable linear $k$-CNF.
\label{theorem-np-complete}
\end{theorem}

\textbf{Proof.} We reduce $k$-SAT to Linear $k$-SAT. Since $k$-SAT is
NP-complete for $k \geq 3$, this will prove the theorem. Let $F$ be a
$k$-CNF. We transform it to a linear $k$-CNF $F'$ such that $F$ is
satisfiable iff $F'$ is. Let $F$ have $m$ clauses and $n$ variables.
For a variable $x$ let $d(x)$ denote the number of times $x$ appears
in $F$. Replace each $x$ by $d(x)$ new variables $x_1, \dots,
x_{d(x)}$.  To ensure that $F$ is satisfiable iff $F'$ is, we force
these variables to take on the same truth value by adding $d(x)$
implication clauses $\{\bar{x}_1, x_2\},
\{\bar{x}_2,x_3\},\dots,\{\bar{x}_{d(x)-1}, x_{d(x)-1}\},
\{\bar{x}_{d(x)}, x_1\}$. Clearly, the new formula $F'$ is linear, and
it is satisfiable iff $F$ is. However, $F'$ is not a $k$-CNF. We
remedy this by adding $k-2$ new variables to each implication clause
and forcing each of them to $0$ by adding a {\em forcer}. A
$\bar{y}$-forcer is a linear $k$-CNF which is satisfiable iff $y$ is set to
$0$. Such a formula can be obtained by taking any minimal
unsatisfiable linear $k$-CNF formula $G_y$ with $y\in\vbl(G)$ and removing
from $G$ all clauses containing $y$. Adding a $\bar{y}$-forcer to $F'$
for each variable $y$ we added to the implication clauses guarantees
that $F'$ is satisfiable iff $F$ is. $F'$
is a linear $k$-CNF, and the proof is complete. $\hfill\Box$

\subsection{Existence of Unsatisfiable Linear $k$-CNFs}

We will complete the NP-completeness proof of Linear $k$-SAT by
showing that unsatisfiable linear $k$-CNFs exist, for any $k \geq 0$.
This answers the main open question from Porschen, Speckenmeyer and
Randerath~\cite{porschen} and establishes the NP-completeness of
Linear $k$-SAT for all $k\geq 3$.

\begin{theorem}
  For any $k \in \mathbb{N}_0$, there are unsatisfiable linear $k$-CNFs.
\label{theorem-existence}
\end{theorem}

\textbf{Proof.} We prove this by induction on $k$. For $k=0$, the
formula $F=\{\{\}\}$ containing only the empty clause is linear and
unsatisfiable.  For the induction step, let $F=\{C_1,\dots,C_m\}$ be
an unsatisfiable linear $k$-CNF. We will construct an unsatisfiable
linear $(k+1)$-CNF formula $F'$. Create $m$ new variables $x_1, \dots,
x_m$. For a clause $D=\{u_1, \dots, u_m\}$ with $u_i \in \{x_i,
\bar{x}_i\}$, define
$$
F \otimes D :=
\left\{ C_i \cup \{u_i\} \ | \ i=1,\dots,m\right\} \ .
$$
$F'$ is a linear $(k+1)$-CNF formula, and every assignment satisfying
$F\otimes D$ satisfies $D$.  Create $2^m$ variable disjoint copies
$F_1,\dots, F_{2^m}$ of $F$, i.e., $\vbl(F_i) \cap \vbl(F_j) =
\emptyset$ for $i \ne j$. By choosing $2^m$ different sign patterns,
we create $2^m$ distinct $m$-clauses $D_1, \dots, D_{2^m}$ over the
variables $x_i$. The formula $\{D_1, \dots, D_{2^m}\}$ is
unsatisfiable. Hence
$$
F' := \bigcup_{i=1}^{2^m} F_i \otimes D_i 
$$
is unsatisfiable, as well. 
Clearly, $F'$ is a linear $(k+1)$-CNF. $\hfill\Box$\\

This proof constitutes an explicit construction, but note the gigantic
growth of the size of the constructed formulas: Let $t(k)$ denote the
number of clauses of the $k$-CNF formula generated in this
construction. Then $t(k+1) = t(k)2^{t(k)}$, so we have $t(1)=2$,
$t(2)=8$, $t(3)=2048$, $t(4)=2048 \times 2^{2048}$.  Fortunately, there is a
much better upper bound, obtained by a probabilistic argument.\\

\begin{theorem}
For every $k \in \mathbb{N}_0$, there exist an unsatisfiable $k$-CNF $F$ with
\begin{displaymath}
|F| \leq k^4 4^{k} \ .
\end{displaymath}
\label{theorem-upper-bound-unsatisfiable-formula}
\end{theorem}

\textbf{Proof.} Fix any $k \in \mathbb{N}_0$. Let $V$ be a set of $n$
variables, $n$ to be specified later. Let $S$ be a linear $k$-set
system over $V$ and write $m := |S|$. From each $s \in S$, build a
$k$-clause by choosing uniformly at random one of the $2^k$ possible
sign patterns.  Do this independently for each $s \in S$ and obtain a
linear $k$-CNF $F$.  Fix an assignment $\alpha$. For every set $s \in
S$, the probability that the clause $C$ built from $s$ is satisfied by
$\alpha$ is $1-2^{-k}$. Since the sign pattern of each clause is
chosen independently, we obtain

$$
\Pr\left( \alpha \textnormal { satisfies } F \right) = \left(1-2^{-k}\right)^m
$$ 

There are $2^n$ different truth assignments to $V$, thus the
probability that at least one of them satisfies $F$ can be estimated
by the union bound:

$$
\Pr \left( F \textnormal{ is satisfiable } \right) \leq 2^n
\left(1-2^{-k}\right)^m
$$

If $2^n \left(1-2^{-k}\right)^m < 1$, then there exists an
unsatisfiable linear $k$-CNF with $m$ clauses and $n$ variables. Since
$1+x < e^x$ for all $x \ne 0$, we have $2^n\left(1-2^{-k}\right)^m <
e^{n\ln 2-m2^{-k}}$, and
\begin{eqnarray}
  e^{n\ln 2-m2^{-k}} 
  & \leq & 1 \Leftrightarrow \nonumber \\
  n\ln 2 - \frac{m}{2^k} & \leq & 0 \Leftrightarrow \nonumber \\
  2^kn\ln 2 & \leq & m \ . \label{bound-n-m}
\end{eqnarray}

That is, if $m \geq 2^kn\ln 2$, then the random formula $F$ is 
unsatisfiable with positive probability.
By Lemma \ref{lower-bound-set-system} we know that there is a linear
$k$-set system $S$ over $n$ elements of size 
\begin{eqnarray}
m = \left\lceil \frac{2n(n-1)}{k^2(k-1)^2} \right\rceil\geq 
\left\lceil\frac{2n^2}{k^4}\right\rceil
\ .
\label{bound-m}
\end{eqnarray} 
Since $m$
grows superlinearly in $n$, we see that for sufficiently large $n$
the last inequality holds, which implies that there is an
unsatisfiable linear $k$-CNF of size $m$ over $n$ variables. To
obtain an upper bound on $n$ and $m$, plug (\ref{bound-m}) into 
(\ref{bound-n-m}):
\begin{eqnarray*}
2^kn\ln 2 & \leq & \frac{2n^2}{k^4}\\
n & \geq & \frac{\ln 2k^4 2^k }{2}
\end{eqnarray*}
Since we are interested in the order of growth for large $k$ rather than
in constant factors, write
$$
m \in \Theta\left( \frac{n^2}{k^4} \right) = \Theta \left( k^4 4^k\right) \ .
$$
Therefore, there is an unsatisfiable linear $k$-CNF $F$ over 
$n \in \Theta\left(k^42^k\right)$ variables having 
$m \in \Theta \left( k^4 4^k\right)$ clauses, and the theorem follows.
$\hfill \Box$\\

This is the best upper bound we have. It is much better than the
explicit construction of Theorem~\ref{theorem-existence}, but it is
still far away from the best lower bound of $\Omega(k2^k)$.

\section{Partial Satisfaction in Linear $2$-CNF Formulas}

It is well known that every unsatisfiable $k$-CNF contains at least
$2^k$ clauses. This bound is tight, since the $k$-CNF $F_k$ consisting
of all $2^k$ clauses over some variable set $V$, $|V| =k$, is
unsatisfiable. Further, for every $k$-CNF, there exists an assignment
satisfying at least $(1 -2^{-k})|F|$ clauses. This can be seen by
choosing a random assignment and calculating the expected number of
satisfied clauses. This bound is also tight, as $F_k$ demonstrates.
This is interesting: The upper bound on the fraction of clauses one can
always satisfy is achieved by a smallest unsatisfiable formula. Since
unsatisfiable {\em linear} $k$-CNFs are much larger than $2^k$, as we
will see, one might suspect that linear $k$-CNFs are more amenable to
partial satisfaction than general $k$-CNFs, i.e., that for at least
some $k$, there is an $r_k > (1-2^{-k})$ such that every linear
$k$-CNF $F$ admits an assignment satisfying $\geq r_k |F|$ of its
clauses. However, this is not true:

\begin{theorem}
  For every $k \in \mathbb{N}$ and $\delta > 0$, there is a linear $k$-CNF
  $F_{k,\delta}$ such that every assignment leaves at least
  fraction of $(1-\delta)2^{-k}$ of all clauses unsatisfied.
\label{theorem-partial-sat}
\end{theorem}

\textbf{Proof.} The proof is similar to the probabilistic proof of
Theorem \ref{theorem-upper-bound-unsatisfiable-formula}: Given $k$,
fix a linear set system $S$ over ground set $V$. Let $n := |V|$, $m :=
|F|$, which will be determined later. Fix an assignment $\alpha$ on
$V$ and build a random formula $F$ over the skeleton $S$ by randomly
choosing the signs of the literals in every clause. Let $F = \{C_1,
\dots, C_m\}$ and define $m$ random variables $X_i$ by
\begin{eqnarray*}
X_i = \left\{
\begin{array}{ll}
0 & \ \textnormal{ if } \alpha \textnormal{ satisfies } C_i, \\
1 & \ \textnormal{ otherwise. } 
\end{array}
\right.
\end{eqnarray*}
Define $X := \sum_{i=1}^m X_i$. Observe that $\mu := \E[X_i] = 2^{-k}$
and $\E[X] = 2^{-k}m$.  We want to bound the probability that less
that $(1-\delta)2^{-k}m$ clauses are unsatisfied by $\alpha$.
First observe that the $X_i$ are independently identically distributed
binary random variables with expectation $2^{-k}$.  Therefore, $X$
has a binomial distribution with expectation
$\mu = 2^{-k}m$, and Chernoff's inequality yields

\begin{eqnarray*}
\Pr\left\{X < (1-\delta)\mu\right\} & < & e^{-\frac{\mu \delta^2}{2}}
\end{eqnarray*}

For a derivation of this inequality see e.g.~\cite{motwani}.  Applying
the union bound, we estimate

\begin{eqnarray*}
\Pr\left\{\exists \alpha \textnormal{ leaving }  \leq 
  (1-\delta)\mu \textnormal{ clauses unsatisfied }\right\} 
 & \leq & 2^n e^{-\frac{\mu \delta^2}{2}} 
\end{eqnarray*}
and want last term to be smaller than $1$. By 
Lemma~\ref{lower-bound-set-system}, we can choose $m \geq \frac{2n^2}{k^4}$
and calculate
\begin{eqnarray}
n \ln 2 - \frac{\mu \delta^2}{2} & < & 0 \nonumber \\
\frac{\mu \delta^2}{2}  = 2^{-k-1}m\delta^2 & > & n \ln 2 \nonumber\\
m & > &2^{k+1} n \ln 2 \delta^{-2} \label{maxsat-ineq} \\
\frac{2n^2}{k^4} & > &2^{k+1} n \ln 2 \delta^{-2} \nonumber \\
n & > & k^42^k \ln 2 \delta^{-2}
\end{eqnarray}
For every fixed $k$ and $\delta > 0$, we can make the last
inequality true by choosing $n$ sufficiently large. Therefore, there
is a positive probability that the randomly chosen formula $F$ does
not have a truth assignment satisfying more than
$(1- (1-\delta)2^{-k})m$ clauses. $\hfill\Box$\\

By setting $\epsilon = \delta2^{-k}$, we see that there is a linear
$k$-CNF $F$ in which no more than $\left(1-2^{-k}+\epsilon\right)|F|$
clauses can be satisfied. Note that the proof of Theorem\
\ref{theorem-partial-sat} is not specific to linear CNFs. For a more
general setting, call a property of formulas {\em structural} if it
only depends on the skeleton of the formula, not on its signs. For a
structural propery $\P$, let $\ex_{\P}(n,k)$ be the maximum number of
clauses a $k$-CNF over $n$ variables having property $\P$ can have.

\begin{theorem}
Let $\P$ be a structural property of CNFs. If for fixed $k$,
$\ex_{\P}(n,k)$ grows superlinearly in $n$, then for every $\epsilon >
0$, there exists a formula $F_{\epsilon}$ for which no truth
assignment $\alpha$ satisfies more than $(1-2^{-k}+\epsilon)
|F_{\epsilon}|$ clauses.
\end{theorem}

\section{Lower Bounds}

After having established that $f(k) \in O (k^42^k)$, we want to obtain
lower bounds on $f(k)$. To be more precise, we show that $f(k) \in
\Omega(k2^k)$. We prove this by repeated application of the Lov\'{a}sz
Local Lemma. For a formula $F$, define the {\em neighborhood} of a
clause $C$ to be 
$$\Gamma(C) := \left\{D \in F \ \big| \ \vbl(D) \cap
  \vbl(C) \ne \emptyset, \ C \ne D \right\} \ .
$$
It follows from the Local Lemma that a $k$-CNF with $|\Gamma(C)| \leq
\frac{1}{4}2^k$ for every clause $C$ is satisfiable (the constant
$\frac{1}{4}$ can be improved upon). Conversely, if $F$ is
unsatisfiable, it contains a clause $C$ with a large neighborhood.  We
find a partial assignment $\alpha$ on $\vbl(C)$ that satisfies $C$ and
a large part of its neighborhood, say at least $c 2^k$ clauses, for
some constant $c$. Since $F$ is linear, applying $\alpha$ deletes at
most one literal from any clause in $\Gamma(C)$, hence $F^{[\alpha]}$
is a $(k-1)$-CNF. Here, we can again apply the Local Lemma and satisfy
$c 2^{k-1}$ clauses, and so on.  Repeating $k$ times, we have
satisfied at least $c (2^k + 2^{k-1} + \dots + 2^1) = c 2^{k+1} - 2c$
clauses.  Unfortunately, this is not enough. We must somehow take
advantage of the fact that though $F^{[\alpha]}$ contains
$(k-1)$-clauses, the neighborhood of a clause in $F^{[\alpha]}$ cannot
contain too many of them.

\begin{lemma}[Lov\'{a}sz Local Lemma]
  Let $A_1, \dots, A_m$ be events in some probability space, and let
  $G$ be a graph with vertices $A_1,\dots, A_m$ and edges $E$ such
  that each $A_i$ is mutually independent of all the events $\left\{
    A_j \ | \ \{A_i, A_j\} \not \in E, \ i \ne j \right\}$. If there
  exist real numbers $0 < \gamma_i < 1$ for $i = 1, \dots, m$
  satisfying

$$
\Pr(A_i) \leq \gamma_i \prod_{j: \{A_i,A_j\} \in E} (1-\gamma_j)
$$

for all $i = 1,\dots,m$, then

$$
\Pr (A_1 \cup A_2 \cup \dots \cup A_m) < 1
$$
\label{lemma-lovasz}
\end{lemma}

For a proof of the Lov\'{a}sz Local Lemma and different versions,
see e.g.~\cite{probmeth}.

\begin{lemma}
  Let $F$ be a CNF not containing any clause of size $\leq 1$.
  If for any $C \in F$ it holds that
$$
\sum_{D \in \Gamma(C)} 2^{-|D|} \leq \frac{1}{4}
$$
then $F$ is satisfiable.
\label{lemma-local}
\end{lemma}

\textbf{Proof.} This is an application of the Lov\'{a}sz Local Lemma.
Let the probability space be the set of all truth assignments to the
$n$ variables in $F$ with the uniform distribution. Write $F=\{C_1,\dots,
C_m\}$ and let $A_i$ be the event that a random
assignments $\alpha$ does not satisfy clause $C_i$.  Let $G$ be the graph
where $A_i$ and $A_j$ are conntected if they have a variable in
common, and let $\gamma_i := 2^{1-|C|} < 1$. For each $i=1,\dots,m$, we have
\begin{eqnarray*}
& & \gamma_i \prod_{j: \{A_i,A_j\} \in E} (1-\gamma_j) \\
& \geq \quad & \gamma_i \left(1 - \sum_{C_j \in \Gamma(C_i)} \gamma_j\right)\\
& = \quad & 2^{1-|C_i|} \left(1 - 2\sum_{C_j \in \Gamma(C_i)}2^{-|D_j|} \right)\\
& \geq \quad & 2^{-|C_i|} = \Pr(A_i)
\end{eqnarray*}
Hence, by Lemma~\ref{lemma-lovasz}, the probability that $\alpha$ leaves
some clause unsatisfied is $< 1$, and thus with positive probability,
$\alpha$ satisfies $F$. Therefore, $F$ is satisfiable.$\hfill\Box$\\

\begin{definition}
  Let $k$ be fixed. An $[l,k]$-CNF is a CNF with $l \leq |C| \leq k$
  for every $C\in F$.  For an $[l,k]$-CNF $F$ and a variable $x$, let
$$
d_F(x) := \left| \left\{ C \in F \ \big| \ x \in \vbl(C),\ |C| \leq
    k-1\right\}\right|
$$ 
If there is no danger of confusion, we will simply write $d(x)$.
Further, let
$$
d(F) := \max_{x \in \vbl(F) }  d_F(x) 
$$
Finally, define
\begin{eqnarray*}
\Gamma'(C) & := & \left\{ D \in \Gamma(C) \ \big| \ |D| \leq k-1\right\} \\
\Gamma'_x(C) & := & \left\{D \in \Gamma'(C) \ \big| \ x \in \vbl(D)\right\}.
\end{eqnarray*}
\end{definition}

\begin{lemma}
Let $F$ be an $[l,k]$-CNF. Then for any clause $C \in F$
$$
|\Gamma'(C)| \leq kd(F)
$$
\label{lemma-shrunk-neighborhood}
\end{lemma}

\textbf{Proof.} We simply calculate
$$
\left|\Gamma'(C)\right| = \left| \bigcup_{x \in \vbl(C)} \Gamma'_x(C) \right|
\leq \sum_{x \in \vbl(C)} \left|\Gamma'_x(C)\right|
\leq |C|d_F(x) \leq kd(F) \ .
$$
and the lemma follows.$\hfill\Box$\\

We need a lemma that states that after setting the variables of $C$
such that $C$ and a large part of its neighborhood is satisfied,
$d(F)$ does not increase too much.

\begin{lemma}
  Let $F$ be a linear $[l,k]$-CNF and $C$ be any clause in $F$. Let
  $\alpha$ be any assignment that is defined only on the variables of
  $C$. If $\alpha$ satisfies $C$, then $d(F^{[\alpha]}) \leq d(F) +
  k$, and $F^{[\alpha]}$ is a linear $[l-1,k]$-CNF.
\end{lemma}

\textbf{Proof.} Since $F$ is linear, $C$ is the only clause containing
more than one variable in the domain of $\alpha$. Since $\alpha$
satisfies $C$, every clauses loses at most one literal, and hence
$F^{[\alpha]}$ is an $[l-1,k]$-CNF. Surely, it is linear as well. To
bound the amount by which $d(F)$ increases, consider any $y \in
\vbl(F)$. If $y$ is set by $\alpha$, then $d_{F^{[\alpha]}}(y)=0$.
Otherwise, $d_{F^{[\alpha]}}(y)$ is at most $d_F(y)$ plus the number
of clauses that count additionally towards $d(y)$, i.e., clauses $D$
with $y \in \vbl(D)$, $|D| = k$ and $|D^{[\alpha]}|=k-1$.  Clearly, $D
\in \Gamma(C)$, otherwise its size would not decrease under $\alpha$.
For each $x\in\vbl(C)$, there is at most one such clause $D \in
\Gamma(C)$, since $x, y \in \vbl(D)$ and $F$ is linear. Hence there
are at most $|C| \leq k$ such clauses, thus $d_{F^{[\alpha]}}(y) \leq
d_F(y)+k$. Therefore, $d(F^{[\alpha]}) \leq d(F)+k$ holds as
well.$\hfill\Box$

\begin{corollary}
  Let $F$ be an unsatisfiable $[l,k]$-CNF for $l \geq 2$. There is a
  partial assignment $\alpha$ such that $F^{[\alpha]}$ is an
  $[l-1,k]$-CNF, $d(F^{[\alpha]}) \leq d(F)+k$, and $\alpha$ satisfies
  at least
$$
\frac{l-1}{2l} \left(2^{k-2} - kd(F)2^{k-l} \right)
$$
clauses of $F$.
\label{corollary-neighborhood}
\end{corollary}

\textbf{Proof.} By Lemma\ \ref{lemma-local}, we know that if $F$ is
unsatisfiable, there is a clause $C \in F$ with
$$
\sum_{D \in \Gamma(C)} 2^{-|D|} > \frac{1}{4}
$$
Further, using Lemma\ \ref{lemma-shrunk-neighborhood}, we can estimate
$$
\sum_{D \in \Gamma(C)} 2^{-|D|} \leq | \Gamma(C)|2^{-k} + kd(F)2^{-l}
$$
And thus, solving for $|\Gamma(C)|$, 
\begin{eqnarray}
|\Gamma(C)| > 2^{k-2} - kd(F)2^{k-l}
\label{ineq-neighborhood}
\end{eqnarray}
Let $x_1, \dots, x_{|C|}$ be the variables of $C$. Since the
$\Gamma_{x_i}(C)$ are pairwise disjoint, by the pigeonhole principle
there is an $x_i$ such that $|\Gamma_{x_1}(C)| \leq |\Gamma(C)| / l$.
Set $\alpha(x_i)$ such that $\alpha$ satisfies $C$. For the remaining
$|C|-1$ variables $x_j$ of $C$ , set $\alpha(x_j)$ such that it
satisfies at least $|\Gamma_{x_j}|/2$. Overall, we satisfy at least
$$
\frac{|C|-1}{2|C|} | \Gamma(C)|
$$
clauses. Inequality (\ref{ineq-neighborhood}) and the fact that $|C| \geq l$ imply the lemma.$\hfill\Box$\\

\begin{theorem}
Let $f(k) := \min \left\{ |F| \ \big| \ F \textnormal{ is an unsatisfiable linear } k\textnormal{-CNF}\right\}$.
Then $f(k) \in \Omega\left(k2^k\right)$.
\end{theorem}

\textbf{Proof.} Let $F$ be an unsatisfiable linear $k$-CNF. We show
that $|F| \in \Omega\left(k2^k\right)$.  Define $F_i$ for $0 \leq i
\leq k-1$ as follows: $F_0 := F$. For $0 \leq i \leq k-2$, apply
Corollary~\ref{corollary-neighborhood} on $F_i$ and let $\alpha_i$ be
a partial assignment as described in the corollary. Define $F_{i+1} =
F_i^{[\alpha_i]}$. It follows that $F_i$ is an unsatisfiable
$[k-i,k]$-CNF, $d(F_i) \leq ik$, and $\alpha_i$ satisfies at least
$$
M(i) := \frac{k-i-1}{2(k-i)} \left(2^{k-2} - k^2i2^i\right)
$$
clauses of $F_i$, i.e., $|F_i| - |F_{i+1}| \geq M(i)$. Hence, for any
$0 \leq j \leq k-1$, we obtain
\begin{eqnarray*}
  |F| & \geq  & \quad \sum_{i=0}^{j-1} M(i) \geq \frac{k-j-1}{2(k-j)} \left( j2^{k-2} -
    k^2\sum_{i=0}^{j-1} i2^i\right) \\
  &  \geq & \quad \frac{k-j-1}{2(k-j)} \left( j2^{k-2} - k^2 j2^j\right) \ .
\end{eqnarray*}
Plugging in for example $j = k - 3\log k$ yields the claimed bound of
$|F| \in \Omega\left(k2^k\right)$. $\hfill\Box$

\section{Small Unsatisfiable Linear $3$-CNFs}

In this section, we construct small unsatisfiable linear $3$-CNFs. However,
we do not know the exact value of $f(3)$. Consider the formula 
$$\{\{\bar{x}_1,x_2\},  \{\bar{x}_2,x_3\},\dots,\{\bar{x}_{n-1},
x_n\}, \{\bar{x}_n, x_1\}\} \ .$$ Every assignment satisfying it sets
all $x_i$ to $1$ or all to $0$. 
We use this formula as a gadget for
building so-called {\em forcers}. A formula $F$ is called a {\em
  $C$-forcer} if every assignment satisfying $F$ satisfies $C$.
Define
$$
F_6 := \{\{\bar{x}_1,x_2\}, \{\bar{x}_2,x_3\}, \{\bar{x}_3,x_4\},
\{\bar{x}_4,x_1\}, \{x_1,x_3\}, \{\bar{x}_2,\bar{x}_4\}
$$
and note that it is unsatisfiable. For a clause $\{u,v,w\}$, define
$$
F_6 (\{u,v,w\}) := \{\{\bar{x}_1,x_2,u\}, \{\bar{x}_2,x_3,v\},
\{\bar{x}_3,x_4,u\}, \{\bar{x}_4,x_1,v\}, \{x_1,x_3,w\},
\{\bar{x}_2,\bar{x}_4,w\}\} \ .
$$
This formula is a linear $3$-CNF and a $\{u,v,w\}$-forcer. A nice
property of this forcer is that no variables in the forced clause
occur together in one of the clauses of a forcer. Taking the union of
the forcers $F_6(C)$ for all $8$ clauses $C$ over $\{u,v,w\}$ (and
renaming the variables $x_i$ each time, to ensure linearity), we
obtain an unsatisfiable linear $3$-CNF with $48$ clauses. This is
exactly the construction used in the proof of
Theorem~\ref{theorem-existence}.\\

We can improve the above construction. Define
\begin{eqnarray*}
F_8 (\{u,v\}) & := & \{\{\bar{x}_1,x_2,u\}, \{\bar{x}_2,x_3,v\},
\{\bar{x}_3,x_4,u\}, \{\bar{x}_4,x_5,v\}, \{\bar{x}_5,x_6,u\}, 
 \{\bar{x}_6,x_1,v\}, \\
& & \{x_1,x_3,x_5\},\{\bar{x}_2,\bar{x}_4,\bar{x}_6\} \} \ .
\end{eqnarray*}
This is ax $\{u,v\}$-forcer with $8$ clauses. Building $4$ forcers
for the clauses $\{u,v\},\{u,\bar{v}\},\{\bar{u},v\},\{\bar{u},\bar{v}\}$,
we obtain an unsatisfiable linear $3$-CNF with $32$ clauses.\\

We go on: Consider $\{\{u,v\},\{u,\bar{v}\},\{\bar{u},v\},
\{\bar{u},\bar{v},w\}, \{\bar{u},\bar{v},\bar{w}\}\}$. Clearly, this formula
is unsatisfiable. Build the forcer $F_8(C)$ for the three $2$-clauses.
Then build $F_6(\{\bar{u},\bar{v},w\})$. Finally, add 
$\{\bar{u},\bar{v},\bar{w}\}\}$ and obtain an unsatisfiable linear $3$-CNF
with $3 \times 8+6+1 = 31$ clauses.\\

The trick here was that we can afford to enforce one clause $C$ not by 
using $F_6(C)$, but by directly including it into our final formula.
Of course, we must keep that final formula linear, hence we cannot apply
this trick too often. However, we can tweak the formula such that we 
can apply this trick twice. Consider
\begin{eqnarray*}
  F_w & = &  \{ \{u,v,w\}, \{\bar{u},v,w\}, \{\bar{v},w\}\} = \{C_1,C_2,C_3\}\\
  F_{\bar{w}} & = & \{ \{\bar{w},x,y\}, \{\bar{w},x,\bar{y}\}, 
  \{\bar{w},\bar{x}\}\}  = \{D_1,D_2,D_3\} \ .
\end{eqnarray*}
The formula $F_w$ is a $\{w\}$-forcer, and $F_{\bar{w}}$ is a
$\{\bar{w}\}$-forcer. We can build a linear formula from which $F_w$
and $F_{\bar{w}}$ can be derived:
$$
F := F_6(C_2) \cup F_6(D_2) \cup F_8(C_3) \cup F_8(D_3) \cup \{C_1\} 
\cup \{D_1\}
$$
Here, we applied the above trick of directly including a desired clause
into the final formula twice, namely to $C_1$ and $D_1$. This is 
an unsatisfiable linear $3$-CNF with $6+6+8+8+1+1 = 30$ clauses.

\section{Conclusion}

We showed that the size of a smallest unsatisfiable linear $k$-CNF is
in $\Omega\left(k2^k\right) \cap O\left(k^4 4^k\right)$. However, the
best constructive upper bound is a tower-like function. It is
desirable to find a way to {\em construct} unsatisfiable $k$-CNFs of
reasonable size, since this will give much better insight into the
structure of those formulas than a probabilistic proof.\\

One approach would be to use hypergraph vertex coloring problems: If
one translates the $k$-colorability problem of a $d$-uniform linear
hypergraph into a CNF in the natural way, one obtains a linear CNF
with clauses of size $k$ and $d$. There are linear $d$-uniform
hypergraphs with arbitrarily large chromatic number, for any $d$. This
follows e.g. from the Hales-Jewett-Theorem~\cite{hales-jewett} on
combinatorial lines. However, the bounds obtained from this theorem
are tower-like, too.

\bibliographystyle{splncs}
\bibliography{refs}

\end{document}